\documentclass[superscriptaddress,twocolumn,10pt,showpacs]{revtex4}
\usepackage{amsfonts,graphicx,amsmath,subfigure,longtable,amssymb,mathrsfs}
\makeatletter

\newcommand{\Rmnum}[1]{\expandafter\@slowromancap\romannumeral #1@}
\makeatother
\begin{document}

\title[Short Title]{Preparation of three-dimensional entanglement for distant atoms in coupled cavities via atomic spontaneous emission and cavity decay}
\author{Shi-Lei Su}
\affiliation{Department of Physics, Harbin Institute of Technology,
Harbin 150001, China}
\author{Xiao-Qiang Shao}
\affiliation{School of Physics, Northeast Normal University, Changchun 130024, China}
\affiliation{Centre for Quantum Technologies, National University of Singapore, 3 Science Drive 2, Singapore 117543}
\author{Hong-Fu Wang}
\affiliation{Department of Physics, College
of Science, Yanbian University, Yanji, Jilin 133002, China}
\author{Shou Zhang\footnote{szhang@ybu.edu.cn}}
\affiliation{Department of Physics, Harbin Institute of Technology,
Harbin 150001, China} \affiliation{Department of Physics, College
of Science, Yanbian University, Yanji, Jilin 133002, China}
\begin{abstract}
We propose a dissipative scheme to prepare a three-dimensional entangled state
for two atoms trapped in separate coupled cavities. Our work shows that both atomic spontaneous emission and cavity decay, which are two typical obstacles in unitary-dynamics-based schemes, could be utilized as resources for high-dimensional entangled state preparation without specifying initial state and controlling time precisely. Final numerical simulation with one group of experimental parameters indicates that the performance of our scheme is better than the unitary-dynamics-based scheme.

 %{\bf{Keywords:}}

\end{abstract}
\maketitle

\section{Introduction}

It is well known that the dissipation induced by the environment is inevitable in the
development of quantum science and technology. For a long time, dissipation has been regarded
as a major obstacle for developing quantum information technology.
Generally, there are two common methods to deal
with the decoherence, one is quantum error
correction~\cite{001,002,003}, which relies on high-fidelity gates
for detecting as well as correcting errors, and the other is to
encode the qubits into a decoherence-free
subspace~(DFS)~in multipartite systems~\cite{004,005,006,007} by
utilizing the certain coupling symmetry between system and
environment. Fundamentally different with the former methods, using
dissipation as powerful resource has special merits since it is used to
create entanglement rather than destroy entanglement~\cite{008,009,010,011,012,013,014,015,016,017,018,b019}.
Particularly, Kastoryano \emph{et al}. consider a
dissipative scheme for preparing a maximally entangled state
of two $\Lambda$ atoms in a high finesse optical cavity without requirement of
state initialization~\cite{011}. Dalla Torre \emph{et al}.
realized the spin squeezing in a dissipative atom-cavity system~\cite{014}.
Leghtas \emph{et al}. also prepared a maximally entangled state of a pair
of superconducting qubits in a low-\emph{Q} cavity. These schemes shows that cavity decay is
no longer undesirable, but plays positive role for state preparation. Nevertheless, the spontaneous emission
plays negative role. Recently, Shao \emph{et al}. proposed a dissipative scheme which shows that for high-dimensional entanglement preparation, the
situation is quite reverse, i.e., spontaneous emission plays effective role rather than cavity decay~\cite{018}.

Coupled cavity model provides an essential tool for distributed quantum
information processing and has been studied both
theoretically~\cite{bb019,019,020,021,022,023,024,025} and experimentally~\cite{026}.
Most of the coupled-cavity-system-based scheme focus on the coherent unitary dynamics that requires
time control and state initialization. Motivated by Ref.~\cite{011}, Shen \emph{et al}. designed a dissipative
scheme to prepare steady-state entanglement in coupled cavities which requires neither definite initial states
nor precise time control~\cite{013}.

High-dimensional entangled states have attracted more and more attentions owing to the fact that they can
enhance the security of quantum key distribution~\cite{027,028} and violate the local realism more strongly
than the two-dimensional
entanglement~\cite{029}. And how to realize high-dimensional entanglement has been researched in the
fields of linear optics experimentally
by utilizing the spatial modes of photons carrying orbital angular momentum information~\cite{030,031}
and of cavity quantum electrodynamics~(QED) theoretically through
the unitary dynamics~\cite{032,033,034,035}.

As is well known to us, atomic spontaneous emission and cavity decay are two typical decoherence factors, which would decrease the
feasibility of the unitary-dynamics-based scheme. The previous works show that dissipative schemes could use
either cavity decay or spontaneous emission alone to prepare entanglement,
but when one of the factors exerts positive effects on state preparation,
the other is just opposite. Thus, using both decoherence factors to
prepare entanglement has unique characteristic. Although cooling schemes meet this goal, more classical fields
are required to resonantly drive the undesired state to single-excitation subspace which would decay
to the desired state probably~\cite{036,037}. In this paper, we propose a dissipative scheme, which makes full use of unitary dynamics provided by microwave field
and dissipative factor originating from spontaneous emission and cavity decay, to prepare
three-dimensional entangled state in coupled cavities. In order to know more clearly about the effect of each dissipative factors, we first consider the system without cavity decay, and then consider it without atomic spontaneous emission.
The analytical and numerical results show that
both cavity decay and atomic spontaneous emission are capable of being useful resources for entanglement preparation. However, the cavity-decay-based case is not as ideal as the spontaneous-emission-based case, which could be improved through adding feedback control. Interestingly, conditions for achieving the effective dissipative channels of spontaneous-emission-based case are almost the same to cavity-decay-based case, which could be satisfied at the same time. Therefore, it is possible to use spontaneous emission and cavity decay simultaneously for state preparation.
There are several main characteristics of our scheme. (i) Our scheme is independent of initial state and do not require precise time control. (ii) Both spontaneous emission and cavity decay are treated as resources in our scheme. (iii) Performance of our scheme is better than unitary-dynamics-based scheme.

The structure of our paper is as follows. In Sec.~\ref{s002}, we briefly introduce the basic model of our scheme and review the effective operators formula.
In Sec.~\ref{s003}, we investigate the effect of spontaneous emission and cavity decay on three-dimensional entangled state generation, respectively. Also, feedback control is added to improve the performance of cavity-decay-based case. Moreover, we study the case that simultaneously utilize spontaneous emission and cavity decay as resources to prepare the desired state. Discussion and conclusion are given out in Sec.~\ref{s004} and Sec.~\ref{s005}, respectively.

\section{Basic Model and Method}\label{s002}

\begin{figure}
\begin{center}
  % Requires \usepackage{graphicx}
  \includegraphics[scale=0.65]{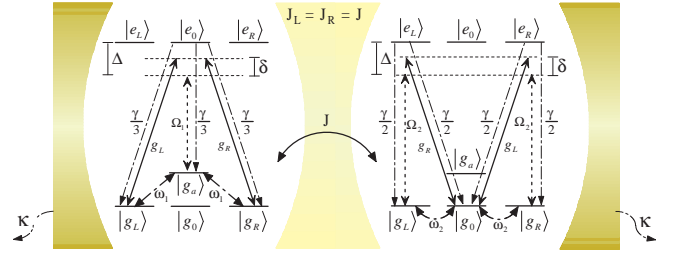}\\
  \caption{Setup for dissipative preparation of three-dimensional entangled state between two $^{87}$Rb atoms trapped in two
  bimode coupled cavities. States $|g_{L}\rangle$, $|g_{0}\rangle$, $|g_{R}\rangle$, and $|g_{a}\rangle$
  correspond to atomic levels $|F = 1, m_{f} = -1\rangle$,
  $|F = 1, m_{f} = 0\rangle$, $|F = 1, m_{f} = 1\rangle$ and $|F = 2, m_{f} = 0\rangle$
  of 5$S_{1/2}$, respectively. And $|e_{L}\rangle$, $|e_{0}\rangle$ and $|e_{R}\rangle$
  correspond to atomic levels $|F = 1, m_{f} = -1\rangle$, $|F = 1, m_{f} = 0\rangle$ and $|F = 1, m_{f} = 1\rangle$ of 5$P_{3/2}$, respectively.}\label{f001}
\end{center}
\end{figure}

Considering a system composed of two $^{87}$Rb atoms trapped in bimode coupled cavities, as shown in Fig.~\ref{f001}.
For the first~(second)~atom, an off-resonance $\pi$-polarized optical laser with detuning $\Delta$, Rabi frequency $\Omega_{1(2)}$ is applied to drive the transition
$|e_{0}\rangle\leftrightarrow |g_{a}\rangle$~($|e_{L}\rangle\leftrightarrow |g_{L}\rangle$ and $|e_{R}\rangle\leftrightarrow |g_{R}\rangle$, respectively). The cavity modes $a_{L1(L2)}$ and $a_{R1(R2)}$ are coupled to the transitions
$|e_{0(R)}\rangle\leftrightarrow |g_{L(0)}\rangle$ and $|e_{0(L)}\rangle\leftrightarrow |g_{R(0)}\rangle$ with detuning $\Delta-\delta$, coupling strength
$g_{L}$ and $g_{R}$, respectively. And a microwave field with Rabi frequency $\omega_{1(2)}$ is introduced to resonantly coupled to
the transition $|g_{L}\rangle\leftrightarrow |g_{a(0)}\rangle$ and $|g_{R}\rangle\leftrightarrow |g_{a(0)}\rangle$, respectively.
Under the rotating-wave approximation, the whole system Hamiltonian in a rotating frame reads
$H = H_{0} + H_{g} + V_{+} + V_{-}$, where

\begin{eqnarray}\label{e001}
% \nonumber to remove numbering (before each equation)
  H_{0} &=& \delta(\hat{a}_{L1}^{\dag}\hat{a}_{L1} + \hat{a}_{L2}^{\dag}\hat{a}_{L2} + \hat{a}_{R1}^{\dag}\hat{a}_{R1} + \hat{a}_{R2}^{\dag}\hat{a}_{R2})
     \cr&& + [g_{L}|g_{L}\rangle_{11}\langle e_{0}|\hat{a}_{L1}^{\dag} + g_{L}|g_{0}\rangle_{22}\langle e_{R}|\hat{a}_{L2}^{\dag}
     \cr&& + g_{R}|g_{R}\rangle_{11}\langle e_{0}|\hat{a}_{R1}^{\dag} + g_{R}|g_{0}\rangle_{22}\langle e_{L}|\hat{a}_{R2}^{\dag} + {\rm H.c.}]
     \cr&& + \Delta(|e_{0}\rangle_{11}\langle e_{0}| + |e_{L}\rangle_{22}\langle e_{L}| + |e_{R}\rangle_{22}\langle e_{R}|)
     \cr&& + J_{L}(\hat{a}_{L1}^{\dag}\hat{a}_{L2} + {\rm H.c.}) + J_{R}(\hat{a}_{R1}^{\dag}\hat{a}_{R2} + {\rm H.c.}), \\
  H_{g} &=& \omega_{1}(|g_{L}\rangle_{11}\langle g_{a}| + |g_{R}\rangle_{11}\langle g_{a}| + {\rm H.c.})~~~~~~~~~~~~~~
      \cr&& + \omega_{2}(|g_{L}\rangle_{22}\langle g_{0}| + |g_{R}\rangle_{22}\langle g_{0}| + {\rm H.c.}),\\
  V_{+} &=& \Omega_{1}(|e_{0}\rangle_{11}\langle g_{a}|) + \Omega_{2}(|e_{L}\rangle_{22}\langle g_{L}| + |e_{R}\rangle_{22}\langle g_{R}|),\\
   V_{-} &=&V_{+}^{\dag},
      \end{eqnarray}
in which $a_{Li}$ and $a_{Ri}$ are the cavity operators in cavity \emph{i} (\emph{i} = 1, 2). $J_{L(R)}$ denotes the photon-hopping strength between two coupled cavities. By introducing four delocalized bosonic modes $\hat{c}_{L1} = (\hat{a}_{L1} - \hat{a}_{L2})/\sqrt{2}$, $\hat{c}_{L2} = (\hat{a}_{L1} + \hat{a}_{L2})/\sqrt{2}$, $\hat{c}_{R1} = (\hat{a}_{R1} - \hat{a}_{R2})/\sqrt{2}$, $\hat{c}_{R2} = (\hat{a}_{R1} + \hat{a}_{R2})/\sqrt{2}$, the Hamiltonian $H_{0}$ can be rewritten as
\begin{eqnarray}\label{e002}
 H_{0} &=& (\delta - J_{L})\hat{c}_{L1}^{\dag}\hat{c}_{L1} + (\delta + J_{L})\hat{c}_{L2}^{\dag}\hat{c}_{L2}
  \cr&& + (\delta - J_{R})\hat{c}_{R1}^{\dag}\hat{c}_{R1} + (\delta + J_{R})\hat{c}_{R2}^{\dag}\hat{c}_{R2}
  \cr&& + \frac{g_{L}}{\sqrt{2}}[|g_{L}\rangle_{11}\langle e_{0}|(\hat{c}_{L1}^{\dag} + \hat{c}_{L2}^{\dag})
  \cr&& + |g_{0}\rangle_{22}\langle e_{R}|(\hat{c}_{L2}^{\dag} - \hat{c}_{L1}^{\dag}) + {\rm H.c.}]
  \cr&& + \frac{g_{R}}{\sqrt{2}}[|g_{R}\rangle_{11}\langle e_{0}|(\hat{c}_{R1}^{\dag} + \hat{c}_{R2}^{\dag})
  \cr&& + |g_{0}\rangle_{22}\langle e_{L}|(\hat{c}_{R2}^{\dag} - \hat{c}_{R1}^{\dag}) + {\rm H.c.}]
  \cr&& + \Delta(|e_{0}\rangle_{11}\langle e_{0}| + |e_{L}\rangle_{22}\langle e_{L}| + |e_{R}\rangle_{22}\langle e_{R}|).
\end{eqnarray}
For simplicity, we set $g_{L} = g_{R} = g$, $\Omega_{1} = \Omega_{2} = \Omega$, and $\omega_{1} = -\omega_{2} = \omega$ in the following.
The photon decay rate of cavity \emph{i} is denoted as $\kappa_{i}$ (\emph{i} = 1, 2)~(suppose two field modes in the same cavity have the same decay rate).
The excited state of the first atom $|e_{0}\rangle$ spontaneously decay into ground states with branching rate $\gamma_{1}/3$, while the state $|e_{L(R)}\rangle$
of the second atom are translated into $|g_{L(R)}\rangle$ and $|g_{0}\rangle$ with rate $\gamma_{2}/2$. We assume $\kappa_{1} = \kappa_{2} = \kappa$ and $\gamma_{1} = \gamma_{2} = \gamma$ throughout this paper. Thus, the Lindblad operators associated with the cavity decay and spontaneous emission can be expressed as
$L^{\kappa,c_{L1}} = \sqrt{\kappa}\hat{c}_{L1}$, $L^{\kappa, c_{R1}} = \sqrt{\kappa}\hat{c}_{R1}$, $L^{\kappa, c_{L2}} = \sqrt{\kappa}\hat{c}_{L2}$, $L^{\kappa, c_{R2}} = \sqrt{\kappa}\hat{c}_{R2}$,
$L^{\gamma_{1},g_{L(a, R)}} = \sqrt{\gamma/3}|g_{L(a, R)}\rangle_{11}\langle e_{0}|$, $L^{\gamma_{2},g_{L(0)}} = \sqrt{\gamma/2}|g_{L(0)}\rangle_{22}\langle e_{L}|$
and $L^{\gamma_{2},g_{R(0)}} = \sqrt{\gamma/2}|g_{R(0)}\rangle_{22}\langle e_{R}|$.
Then, the dynamics of our system is governed by the master equation
\begin{equation}\label{e003}
\dot{\rho} = -i[H, \rho] + \sum_{j}\big [L^{j}\rho L^{j\dag} - \frac{1}{2}( L^{j\dag}L^{j}\rho + \rho L^{j\dag}L^{j})\big].
\end{equation}
Under the condition that the Rabi frequency $\Omega$ of the optical pumping laser is sufficiently weak, the excited states of the atoms and the cavity field modes can be adiabatically eliminated when the excited states are not initially populated. In this case, according to the effective operator method in Ref.~\cite{038},
we can get the effective master equation as
\begin{equation}\label{e004}
\dot{\rho} = -i[H_{\rm eff}, \rho] + \sum_{j}\big [L^{j}_{\rm eff}\rho L^{j\dag}_{\rm eff} - \frac{1}{2}( L^{j\dag}_{\rm eff}L^{j}_{\rm eff}\rho + \rho L^{j\dag}_{\rm eff}L^{j}_{\rm eff})\big],
\end{equation}
where
\begin{eqnarray}\label{e005}
H_{\rm eff} &=& -\frac{1}{2}[V_{-}H_{\rm NH}^{-1}V_{+} + V_{-}(H_{\rm NH}^{-1})^{\dag}V_{+}] + H_{g},
\cr L^{j}_{\rm eff}&=&L^{j}H_{\rm NH}^{-1}V_{+}.
\end{eqnarray}
In Eq.~(\ref{e005}), $H_{\rm NH}=H_{0} - \frac{i}{2}\sum_{j}L^{j\dag}L^{j}$ is a non-Hermitian Hamiltonian, and its inverted matrix is $H_{\rm NH}^{-1}$.

In the following text, we use the effective operator method to simplify the system and research the dissipative process. Nevertheless, for the sake of preciseness, full hamiltonian \emph{H} rather than $H_{\rm eff}$ is used for numerical simulation to assess the performance of this scheme.

\section{Dissipative preparation of three-dimensional entanglement}\label{s003}
\subsection{Use spontaneous emission as resource}
\begin{figure}
  \begin{center}
    \subfigure[]{\label{f002a}\includegraphics[scale=0.5]{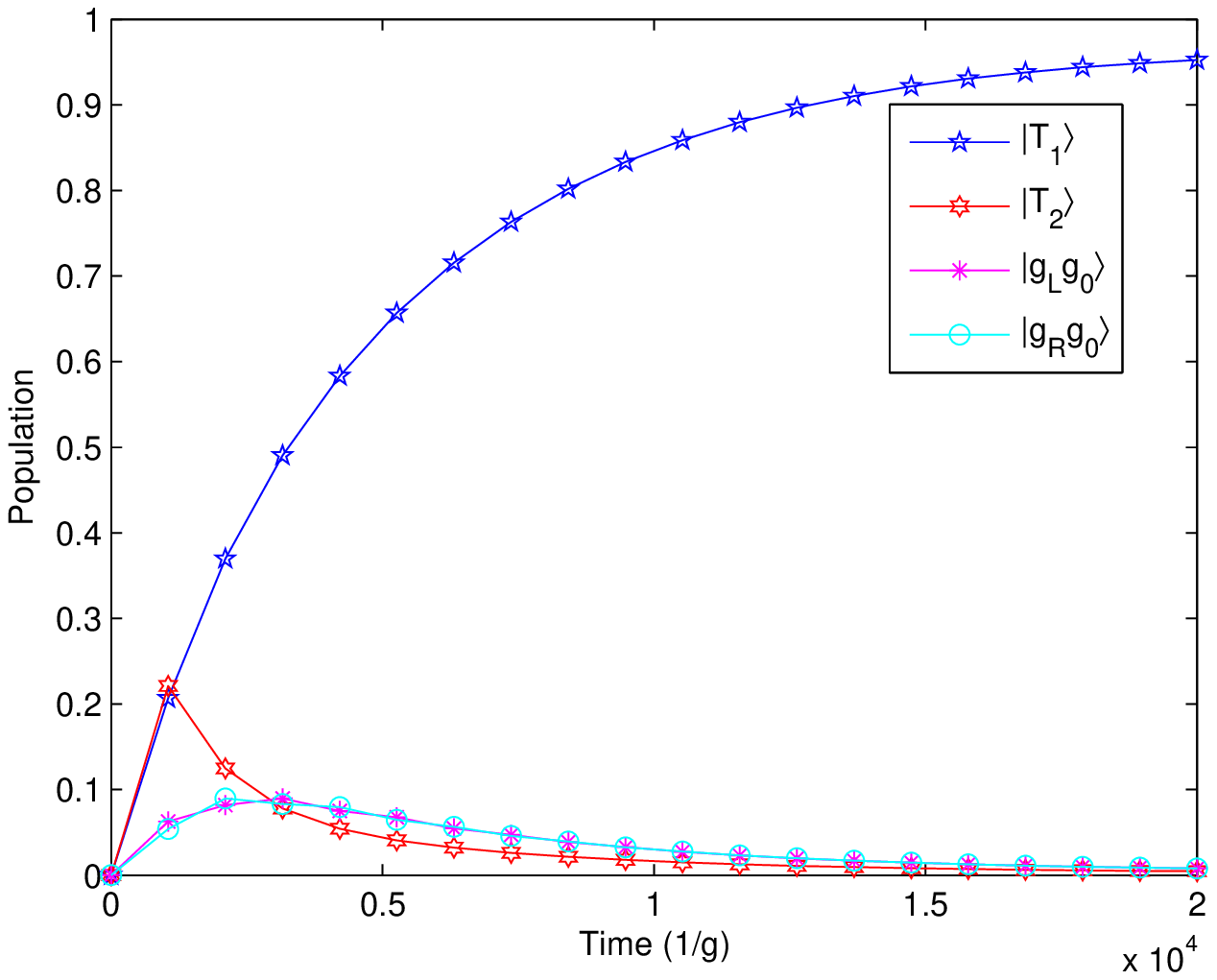}}
    \subfigure[]{\label{f002b}\includegraphics[scale=0.5]{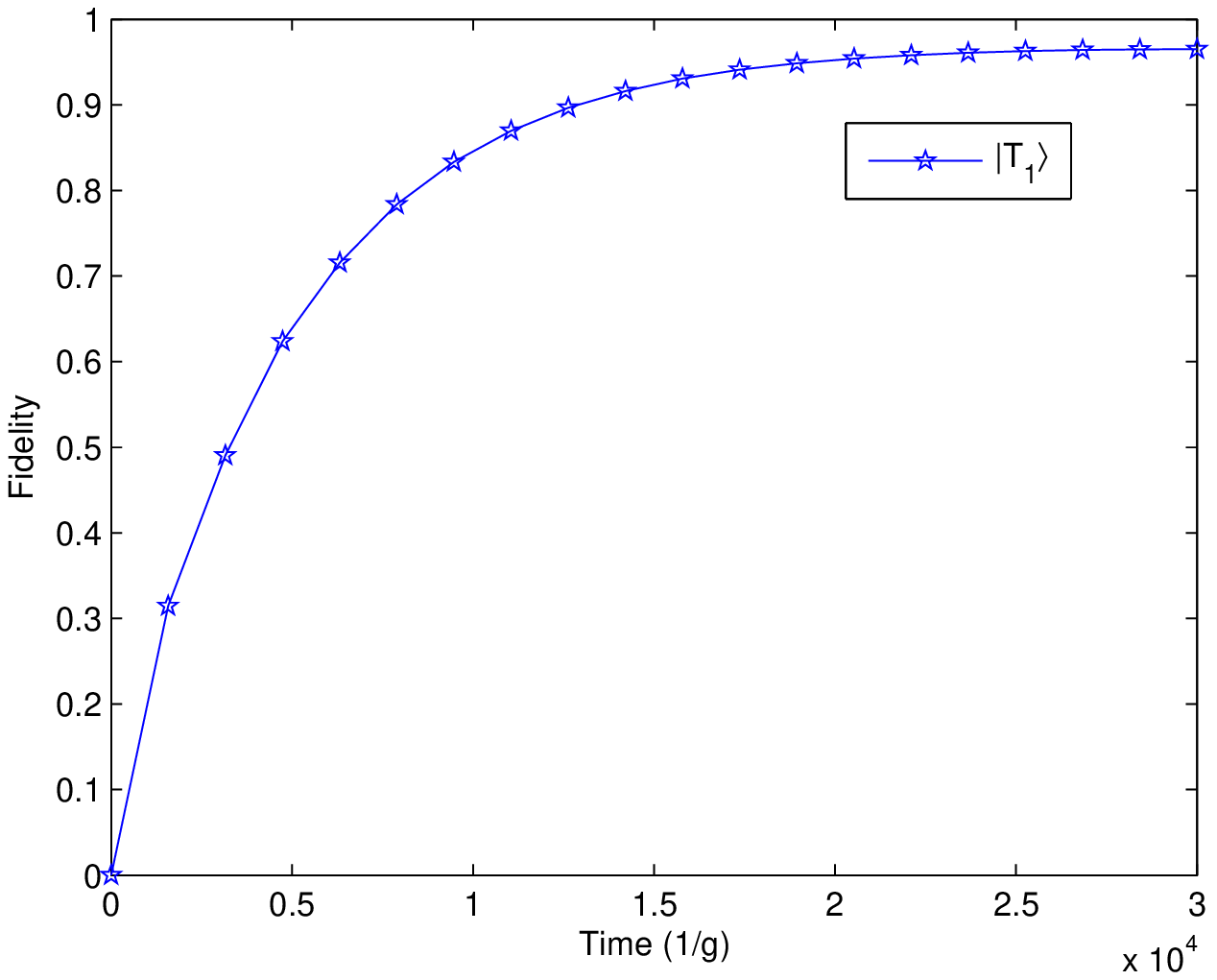}} \\
   \end{center}
\caption{(a) Population for the states in the steady subspace of the spontaneous-emission-based case from an initial state $|g_{a}g_{L}\rangle$.
(b) Fidelity of $|T_{1}\rangle$ state from an initial state $|g_{a}g_{L}\rangle$.Both figures are plotted under the given parameters $\Omega = 0.01g, \omega = 0.2\Omega, \gamma = 0.04g, \kappa = 0, \Delta = g, J = 6g$.}
  \label{f002}
\end{figure}

In this subsection, aiming to gain better insight into the effect of spontaneous emission
on the preparation of an three-dimensional entanglement, we first consider a perfect cavity without decay.
According to Eq.~(\ref{e005}), we have the effective Hamiltonian
\begin{eqnarray}\label{e006}
    H_{\rm eff}& =& \Omega^{2}{\rm Re}[\frac{-\widetilde{J^2}}{g^2\delta + \widetilde{J^2} \widetilde{\Delta}}](|g_{L}g_{L}\rangle\langle g_{L}g_{L}|
   + |g_{R}g_{R}\rangle\langle g_{R}g_{R}|
   \cr& +&  |T_{3}\rangle\langle T_{3}|) + \Omega^{2}{\rm Re}[\frac{-\widetilde{J^{2}}}{2g^2\delta + \widetilde{J^{2}}\widetilde{\Delta}} + \frac{-\widetilde{J^{2}}}{g^2\delta + \widetilde{J^{2}}\widetilde{\Delta}}]
   \cr&\times&(|g_{a}g_{L}\rangle\langle g_{a}g_{L}| + |g_{a}g_{R}\rangle\langle g_{a}g_{R}|)
   \cr& +&  \frac{\Omega^2}{3}{\rm Re}[\frac{g^2(4J + 5\delta) + 3\widetilde{J^2}\widetilde{\Delta}}{2g^4 - 3g^2\delta\widetilde{\Delta} - \widetilde{J^2}\widetilde{\Delta}^2}]|T_{1}\rangle\langle T_{1}|
   \cr& +&  \frac{\Omega^2}{3}{\rm Re}[\frac{-4g^2(J-\delta) + 3\widetilde{J^2}\widetilde{\Delta}}{2g^4 - 3g^2\delta\widetilde{\Delta} - \widetilde{J^2}\widetilde{\Delta}^2}]|T_{2}\rangle\langle T_{2}|
   \cr& +&  \frac{\sqrt{2}\Omega^2}{3}{\rm Re}[\frac{-g^2(J-\delta)}{2g^4 - 3g^2\delta\widetilde{\Delta} - \widetilde{J^2}\widetilde{\Delta}^2}](|T_{1}\rangle\langle T_{2}|
   \cr& +&|T_{2}\rangle\langle T_{1}|) + H_{g},
\end{eqnarray}
in which
\begin{eqnarray}\label{e007}
% \nonumber to remove numbering (before each equation)
  |T_{1}\rangle &=& \frac{1}{\sqrt{3}}(|g_{L}g_{R}\rangle + |g_{R}g_{L}\rangle + |g_{a}g_{0}\rangle),
  \cr|T_{2}\rangle &=& \frac{1}{\sqrt{6}}(|g_{L}g_{R}\rangle + |g_{R}g_{L}\rangle - 2|g_{a}g_{0}\rangle),
  \cr|T_{3}\rangle &=& \frac{1}{\sqrt{2}}(|g_{L}g_{R}\rangle - |g_{R}g_{L}\rangle),
  \cr\widetilde{J^2} &=&J^2-\delta^2,
  \cr\widetilde{\Delta} &=&\Delta - \frac{i\gamma}{2}.
\end{eqnarray}
And $|T_{1}\rangle$ is the desired three-dimensional entangled state.
In addition, on the basis of Eq.~(\ref{e005}), the effective Lindblad operators induced by spontaneous emission are
\begin{eqnarray}\label{e008}
% \nonumber to remove numbering (before each equation)
   L^{\gamma_{1},g_{L(a,R)}}_{\rm eff}&=&\sqrt{\frac{\gamma}{3}}\frac{\Omega \widetilde{J^2}}{2g^2\delta + \widetilde{J^2}\widetilde{\Delta}}
   (|g_{L(a,R)}g_{L}\rangle\langle g_{a}g_{L}|
  \cr& + & |g_{L(a,R)}g_{R}\rangle\langle g_{a}g_{R}|)
   \cr& + & \frac{\sqrt{\gamma}\Omega}{3}[\frac{-g^2(2J+\delta)-\widetilde{J^2}\widetilde{\Delta}}{2g^4-3g^2\delta\widetilde{\Delta}-\widetilde{J^2}\widetilde{\Delta}^2}]
   |g_{L(a,R)}g_{0}\rangle\langle T_{1}|
\cr& + & \frac{\sqrt{2\gamma}\Omega}{3}[\frac{-g^2(J-\delta)+\widetilde{J^2}\widetilde{\Delta}}{2g^4-3g^2\delta\widetilde{\Delta}-\widetilde{J^2}\widetilde{\Delta}^2}]
   |g_{L(a,R)}g_{0}\rangle\langle T_{2}|
\cr L^{\gamma_{2},g_{L(R)}}_{\rm eff}&=&\frac{\sqrt{6\gamma}\Omega}{6}[\frac{-g^2(J+2\delta)-\widetilde{J^2}\widetilde{\Delta}}{2g^4-3g^2\delta\widetilde{\Delta}-\widetilde{J^2}\widetilde{\Delta}^2}]
   |g_{R(L)}g_{L(R)}\rangle\langle T_{1}|
\cr& + &
\frac{\sqrt{3\gamma}\Omega}{6}[\frac{2g^2(J-\delta)-\widetilde{J^2}\widetilde{\Delta}}{2g^4-3g^2\delta\widetilde{\Delta}-\widetilde{J^2}\widetilde{\Delta}^2}]
   |g_{R(L)}g_{L(R)}\rangle\langle T_{2}|
\cr& + &
\frac{\sqrt{2\gamma}\Omega}{2}\frac{\widetilde{J^2}}{g^2\delta + \widetilde{J^2}\widetilde{\Delta}}(|g_{L(R)}g_{L(R)}\rangle\langle g_{L(R)}g_{L(R)}|
\cr& + & |g_{a}g_{L(R)}\rangle\langle g_{a}g_{L(R)}| -(+) \frac{1}{\sqrt{2}}|g_{R(L)}g_{L(R)}\rangle\langle T_{3}|)
\cr L^{\gamma_{2},g_{0}}_{\rm eff}&=&\frac{\sqrt{2\gamma}\Omega}{2}\frac{\widetilde{J^2}}{g^2\delta + \widetilde{J^2}\widetilde{\Delta}}\Big[|g_{L}g_{0}\rangle\langle g_{L}g_{L}| +|g_{a}g_{0}\rangle\langle g_{a}g_{L}|
\cr& + & |g_{a}g_{0}\rangle\langle g_{a}g_{R}| + |g_{R}g_{0}\rangle\langle g_{R}g_{R}|
\cr& + &\frac{1}{\sqrt{2}}(|g_{L}g_{0}\rangle - |g_{R}g_{0}\rangle)\langle T_{3}|\Big]
\cr& + & \frac{\sqrt{6\gamma}\Omega}{6}[\frac{-g^2(J+2\delta)-\widetilde{J^2}\widetilde{\Delta}}{2g^4-3g^2\delta\widetilde{\Delta}-\widetilde{J^2}\widetilde{\Delta}^2}]
(|g_{L}g_{0}\rangle \cr& + & |g_{R}g_{0}\rangle)\langle T_{1}|
\cr& + &
\frac{\sqrt{3\gamma}\Omega}{6}[\frac{2g^2(J-\delta)-\widetilde{J^2}\widetilde{\Delta}}{2g^4-3g^2\delta\widetilde{\Delta}-\widetilde{J^2}\widetilde{\Delta}^2}]
(|g_{L}g_{0}\rangle \cr& + & |g_{R}g_{0}\rangle)\langle T_{2}|.
\end{eqnarray}
It is important to note that if $\Delta\gg\gamma$, $\delta\Delta\geqslant 2g^2$, and the cavity detuning from two photon resonance $\delta$ satisfies the condition $\delta =
(g^2+\sqrt{g^4+4 J^2 \Delta ^2})/(2\Delta)$, other effective decay channels are approximately ignored except the following dominant parts
\begin{eqnarray}\label{e009}
% \nonumber to remove numbering (before each equation)
L^{\gamma_{2},g_{L(R)}}_{\rm eff}&=&
\frac{\sqrt{2\gamma}\Omega}{2}\frac{\widetilde{J^2}}{g^2\delta + \widetilde{J^2}\widetilde{\Delta}}(|g_{L(R)}g_{L(R)}\rangle\langle g_{L(R)}g_{L(R)}|
\cr& + & |g_{a}g_{L(R)}\rangle\langle g_{a}g_{L(R)}| -(+) \frac{1}{\sqrt{2}}|g_{R(L)}g_{L(R)}\rangle\langle T_{3}|)
\cr L^{\gamma_{2},g_{0}}_{\rm eff}&=&\frac{\sqrt{2\gamma}\Omega}{2}\frac{\widetilde{J^2}}{g^2\delta + \widetilde{J^2}\widetilde{\Delta}}\Big[|g_{L}g_{0}\rangle\langle g_{L}g_{L}| +|g_{a}g_{0}\rangle\langle g_{a}g_{L}|
\cr& + & |g_{a}g_{0}\rangle\langle g_{a}g_{R}| + |g_{R}g_{0}\rangle\langle g_{R}g_{R}|
\cr& + &\frac{1}{\sqrt{2}}(|g_{L}g_{0}\rangle - |g_{R}g_{0}\rangle)\langle T_{3}|\Big].
\end{eqnarray}
Since $|g_{L}g_{R}\rangle$, $|g_{R}g_{L}\rangle$ and $|g_{a}g_{0}\rangle$ can be represented by $|T_{1}\rangle$, $|T_{2}\rangle$ and $|T_{3}\rangle$,
the dissipative dynamics in Eq.~(\ref{e009}) would transfer any initial states into the subspace composed of $|T_{1}\rangle$, $|T_{2}\rangle$, $|g_{L}g_{0}\rangle$ and $|g_{R}g_{0}\rangle$. Besides, coherent dynamics governed by Eq.~(\ref{e006}) can be decomposed into two parts, terms consisting of $\Omega^2$($O(\Omega^2)$) and $H_{g}$. $O(\Omega^2)$ induces the transition $|T_{1}\rangle\leftrightarrow|T_{2}\rangle$ and keeps other states invariant. $H_{g}$ keeps $|T_{1}\rangle$
invariant while makes $|T_{2}\rangle$, $|g_{L}g_{0}\rangle$ and $|g_{R}g_{0}\rangle$ evolve out of the subspace. If $\Omega^2\ll\omega$, the unitary dynamics is mainly governed by $H_{g}$ rather than $O(\Omega^2)$. And the condition $\omega_{1} = -\omega_{2}$ is critical since it guarantees $|T_{1}\rangle$ to be the dark state of $H_{g}$. In Fig.~\ref{f002}, we plot the fidelity of state $|T_{1}\rangle$, $F = \langle T_{1}|\rho|T_{1}\rangle$, with the full Hamiltonian and master equation, from which we can see that the desired state can be achieved with a higher fidelity.

\subsection{Use cavity decay as resource}
In this subsection, aiming to gain better insight into the effect of cavity decay
on the preparation of an three-dimensional entanglement, we do not consider spontaneous emission here.

\subsubsection{without feedback control}

According to Eq.~(\ref{e005}), the effective Hamiltonian is achieved as
\begin{eqnarray}\label{e010}
% \nonumber to remove numbering (before each equation)
  H_{\rm eff}& =& \Omega^2{\rm Re}[\frac{-\widetilde{J^2}^{'}}{g^2\widetilde{\delta} + \Delta\widetilde{J^2}^{'}}]
  (|g_{L}g_{L}\rangle\langle g_{L}g_{L}|
   + |g_{R}g_{R}\rangle\langle g_{R}g_{R}|
   \cr& +&  |T_{3}\rangle\langle T_{3}|) + \Omega^2{\rm Re}[\frac{-\widetilde{J^2}^{'}}{g^2\widetilde{\delta} + \Delta\widetilde{J^2}^{'}}
   + \frac{-\widetilde{J^2}^{'}}{2g^2\widetilde{\delta} + \Delta\widetilde{J^2}^{'}}]
   \cr&\times&(|g_{a}g_{L}\rangle\langle g_{a}g_{L}| + |g_{a}g_{R}\rangle\langle g_{a}g_{R}|)
   \cr& +&  \frac{\Omega^2}{3}{\rm Re}[\frac{g^2(4J + 5\widetilde{\delta}) + 3\widetilde{J^2}^{'}\Delta}{2g^4 - 3g^2\widetilde{\delta}\Delta - \widetilde{J^2}^{'}\Delta^2}]|T_{1}\rangle\langle T_{1}|
\cr& +&  \frac{\Omega^2}{3}{\rm Re}[\frac{-4g^2(J-\widetilde{\delta}) + 3\widetilde{J^2}^{'}\Delta}{2g^4 - 3g^2\widetilde{\delta}\Delta - \widetilde{J^2}^{'}\Delta^2}]|T_{2}\rangle\langle T_{2}|
\cr& +&  \frac{\sqrt{2}\Omega^2}{3}{\rm Re}[\frac{-g^2(J-\widetilde{\delta})}{2g^4 - 3g^2\widetilde{\delta}\Delta - \widetilde{J^2}^{'}\Delta^2}](|T_{1}\rangle\langle T_{2}|
   \cr& +&|T_{2}\rangle\langle T_{1}|) + H_{g},
\end{eqnarray}
in which
\begin{eqnarray}\label{e011}
% \nonumber to remove numbering (before each equation)
  \cr\widetilde{J^2}^{'} &=&J^2-(\widetilde{\delta})^2,
  \cr\widetilde{\delta} &=&\delta - \frac{i\kappa}{2}.
\end{eqnarray}
Besides, the effective Lindblad operators induced by cavity decay can be written as
\begin{eqnarray}\label{e012}
% \nonumber to remove numbering (before each equation)
  L^{\kappa, c_{L1(R1)}}_{\rm eff} &=& \frac{\sqrt{2\kappa}\Omega}{2}\frac{g(J+\widetilde{\delta})}{2g^2\delta+\Delta\widetilde{J^2}^{'}}(|g_{L(R)}g_{L}\rangle\langle g_{a}g_{L}|
\cr &+& |g_{R(L)}g_{R}\rangle\langle g_{a}g_{R}|)
\cr &-&\frac{\sqrt{2\kappa}\Omega}{2}\frac{g(J+\widetilde{\delta})}{g^2\delta+\Delta\widetilde{J^2}^{'}}(|g_{a}g_{0}\rangle\langle g_{a}g_{R(L)}|
\cr &+& |g_{R(L)}g_{0}\rangle\langle g_{R(L)}g_{R(L)}| +(-) \frac{1}{\sqrt{2}}|g_{L(R)}g_{0}\rangle\langle T_{3}|)
\cr &-& \frac{\sqrt{6\kappa}\Omega}{6}\frac{g^3}{2g^4 - 3g^2\widetilde{\delta}\Delta - \widetilde{J^2}^{'}\Delta^2}|g_{L(R)}g_{0}\rangle\langle T_{1}|
\cr &+& \frac{\sqrt{3\kappa}\Omega}{6}\frac{g(-4g^2 + 3\Delta(J+\widetilde{\delta}))}{2g^4 - 3g^2\widetilde{\delta}\Delta - \widetilde{J^2}^{'}\Delta^2}
|g_{L(R)}g_{0}\rangle\langle T_{2}|
\cr L^{\kappa, c_{L2(R2)}}_{\rm eff} &=& -\frac{\sqrt{2\kappa}\Omega}{2}\frac{g(J-\widetilde{\delta})}{2g^2\widetilde{\delta} + \Delta \widetilde{J^2}^{'}}
(|g_{L(R)}g_{L}\rangle\langle g_{a}g_{L}|
\cr &+& |g_{L(R)}g_{R}\rangle\langle g_{a}g_{R}|)
\cr &-& \frac{\sqrt{2\kappa}\Omega}{2}\frac{g(J-\widetilde{\delta})}{g^2\widetilde{\delta} + \Delta \widetilde{J^2}^{'}}(|g_{a}g_{0}\rangle\langle g_{a}g_{R(L)}|
\cr &+& |g_{R(L)}g_{0}\rangle\langle g_{R(L)}g_{R(L)}| +(-) \frac{1}{\sqrt{2}}|g_{L(R)}g_{0}\rangle\langle T_{3}|)
\cr &+& \frac{\sqrt{6\kappa}\Omega}{6}\frac{g(3g^2 + 2\Delta(J-\widetilde{\delta}))}{2g^4 - 3g^2\widetilde{\delta}\Delta - \widetilde{J^2}^{'}\Delta^2}|g_{L(R)}g_{0}\rangle\langle T_{1}|
\cr &-& \frac{\sqrt{3\kappa}\Omega}{6}\frac{g\Delta(J-\widetilde{\delta})}{2g^4 - 3g^2\widetilde{\delta}\Delta - \widetilde{J^2}^{'}\Delta^2}|g_{L(R)}g_{0}\rangle\langle T_{2}|
\end{eqnarray}
It is noticeable if $\delta\gg\kappa$, $\delta\Delta\geqslant 2g^2$, and the cavity detuning from two photon resonance $\delta$ meets the condition $\delta =
(g^2+\sqrt{g^4+4 J^2 \Delta ^2})/(2\Delta)$, other effective decay channels are approximately ignored except the following dominant terms
\begin{eqnarray}\label{e013}
% \nonumber to remove numbering (before each equation)
  L^{\kappa, c_{L1(R1)}}_{\rm eff} &=& -\frac{\sqrt{2\kappa}\Omega}{2}\frac{g(J+\widetilde{\delta})}{g^2\delta+\Delta\widetilde{J^2}^{'}}(|g_{a}g_{0}\rangle\langle g_{a}g_{R(L)}|
\cr &+& |g_{R(L)}g_{0}\rangle\langle g_{R(L)}g_{R(L)}| +(-) \frac{1}{\sqrt{2}}|g_{L(R)}g_{0}\rangle\langle T_{3}|)
\cr L^{\kappa, c_{L2(R2)}}_{\rm eff} &=& -\frac{\sqrt{2\kappa}\Omega}{2}\frac{g(J-\widetilde{\delta})}{g^2\widetilde{\delta} + \Delta \widetilde{J^2}^{'}}(|g_{a}g_{0}\rangle\langle g_{a}g_{R(L)}|
\cr &+& |g_{R(L)}g_{0}\rangle\langle g_{R(L)}g_{R(L)}| \cr &+&(-) \frac{1}{\sqrt{2}}|g_{L(R)}g_{0}\rangle\langle T_{3}|)
\end{eqnarray}
Owing to the fact that $|g_{L}g_{R}\rangle$, $|g_{R}g_{L}\rangle$ and $|g_{a}g_{0}\rangle$ can be represented by $|T_{1}\rangle$, $|T_{2}\rangle$ and $|T_{3}\rangle$,
the dissipative channels in Eq.~(\ref{e013}) results into the subspace composed of $|T_{1}\rangle$, $|T_{2}\rangle$, $|g_{L}g_{0}\rangle$ and $|g_{R}g_{0}\rangle$ for any initial states. And the unitary dynamics induced by $O(\Omega^2)$ and $H_{g}$ guarantee $|T_{1}\rangle$ being invariant while the other three states being driven out of the steady subspace.
\begin{figure}
  \begin{center}
    \subfigure[]{\label{f003a}\includegraphics[scale=0.5]{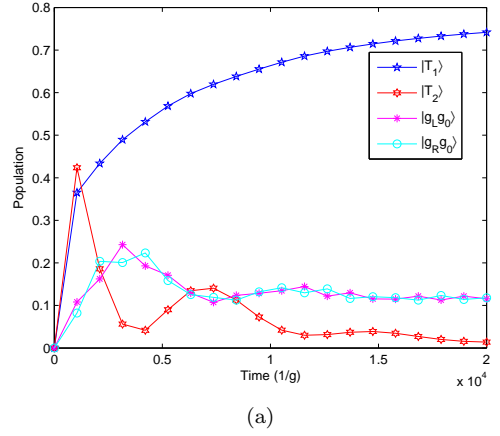}}
    \subfigure[]{\label{f003b}\includegraphics[scale=0.5]{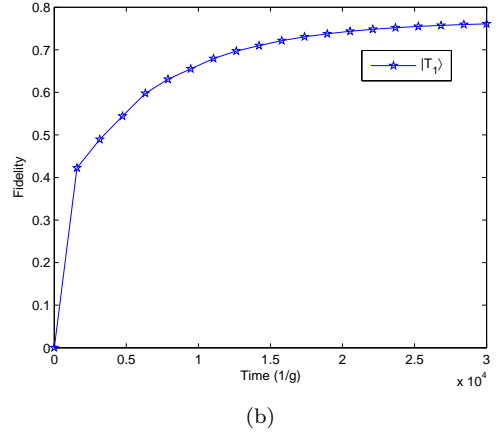}} \\
   \end{center}
\caption{(a) Population for the states in the steady subspace of the cavity-decay-based case from an initial state $|g_{a}g_{L}\rangle$.
(b) Fidelity of $|T_{1}\rangle$ state from an initial state $|g_{a}g_{L}\rangle$. Both figures are plotted under the given parameters $\Omega = 0.03g,
\omega = 0.05\Omega, \gamma = 0, \kappa = 0.05g, \Delta = g, J = 6g$.}
  \label{f003}
\end{figure}
Thus, the desired state could be prepared with any initial state. We plot the population and fidelity in Fig.~\ref{f003} under one group of the optimal parameters. Nevertheless, the results do about 20\% worse than spontaneous-emission-based case because both of state $|g_{L}g_{0}\rangle$ and $|g_{R}g_{0}\rangle$ occupy a population more than 10\% when the system approaches to stabilization. This phenomenon is not hardly to understand through comparing Eq.~(\ref{e013}) with Eq.~(\ref{e009}). Despite that $|g_{L}g_{0}\rangle$ and $|g_{R}g_{0}\rangle$ are driven out of the steady subspace constantly by $H_{g}$, effective decay terms in
Eq.~(\ref{e013}) drives the outside state into states $|g_{L}g_{0}\rangle$ and $|g_{R}g_{0}\rangle$ to some extent. $|g_{L}g_{0}\rangle$ and $|g_{R}g_{0}\rangle$ occupy a population more than 10\% when these two opposite process redress the balance. However, this phenomenon is not exist in the spontaneous-emission-based case since $|g_{R(L)}g_{L(R)}\rangle\langle T_{3}|$ term in Eq.~(\ref{e009}) plays a key role to translate the outside state into $|T_{1}\rangle$ rather than $|g_{L}g_{0}\rangle$ and $|g_{R}g_{0}\rangle$ in some extent.

\subsubsection{improve the performance via feedback control}

In this paragraph, we aim to use the feedback control~\cite{039,040,041,042,043,044,045,046,047,048} to improve the performance of the cavity-decay-based case. The dynamics include feedback control is governed by the master equation
\begin{eqnarray}\label{e014}
% \nonumber to remove numbering (before each equation)
\dot{\rho} &=& -i[H, \rho] + \kappa\sum_{\ell}\big [\hat{U}_{\ell}\hat{\ell}~\rho~\hat{\ell}^{\dag} \hat{U}_{\ell}^{\dag}
- \frac{1}{2}(\ell^{\dag}\ell\rho + \rho\ell^{\dag}\ell)\big],~~~~~
\end{eqnarray}
in which $\ell$ denotes $c_{L1}$, $c_{R1}$, $c_{L2}$ and $c_{R2}$, respectively, and $U_{\ell}$ is the feedback operation.
The main aim for choosing the feedback is to drive $|g_{L}g_{0}\rangle$ and $|g_{R}g_{0}\rangle$ out of the steady state subspace.
Without generality, we choose the feedback operations $\{U_{c_{L1}} = I$, $U_{c_{R1}} = \exp(i\pi\sigma_{x1}/2)$, $U_{c_{L2}} = I$, $U_{c_{R1}}= I\}$
and $\{U_{c_{L1}} = I$, $U_{c_{R1}} = \exp(i\pi\sigma_{x2}/2)$, $U_{c_{L2}} = I$, $U_{c_{R1}}= I\}$, in which $\sigma_{x1(2)} = |g_{0}\rangle_{22}\langle g(e)_{R}| + |g(e)_{R}\rangle_{22}\langle g_{0}|$, to research the effect of feedback control. Figure~4 shows that fidelity of the cavity-decay-based case can be improved via feedback control.
\begin{figure}
  % Requires \usepackage{graphicx}
  \includegraphics[scale=0.5]{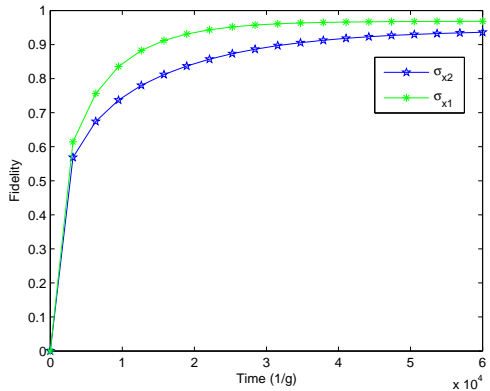}\\
  \caption{Fidelity of cavity-decay-based case via feedback control. The figure is plotted under the parameters $\Omega = 0.04g,
\omega = 0.05\Omega, \gamma = 0, \kappa = 0.1g, \Delta = g, J = 6g$.}\label{f004}
\end{figure}

\subsection{Simultaneously use both spontaneous emission and cavity decay as resources}

Note that the conditions to obtain the effective decay channels in Eq.~(\ref{e009}) and Eq.~(\ref{e013})
can be satisfied at the same time, it is thus possible to use both spontaneous emission and cavity decay to prepare the desired state simultaneously.
In Fig.~5, we plot the fidelity of the presented scheme based on spontaneous emission and cavity decay simultaneously, from which we can see that $|T_{1}\rangle$ can be achieved with the fidelity close to 0.9 under specific parameters.
In Fig.~6, we plot the fidelity with parameters the same to Fig.~5 via added feedback control. Results show that feedback control can improve the fidelity, shorten the time to be steady and improve the robustness on parameters variation.
\begin{figure}
  % Requires \usepackage{graphicx}
  \includegraphics[scale=0.5]{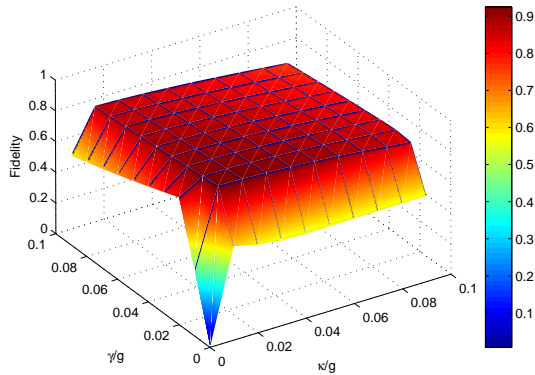}\\
  \caption{Fidelity of the scheme based on spontaneous emission and cavity decay without feedback control at the time 60000/\emph{g}.
  Parameters are chosen as $\Omega = 0.03g, \omega = 0.2\Omega, \Delta = g, J = 6g$. $\kappa$ and $\gamma$ are chosen form 0 to 0.1\emph{g} with regular intervals equal to 0.01\emph{g}.}\label{f005}
\end{figure}
\begin{figure}
  % Requires \usepackage{graphicx}
  \includegraphics[scale=0.5]{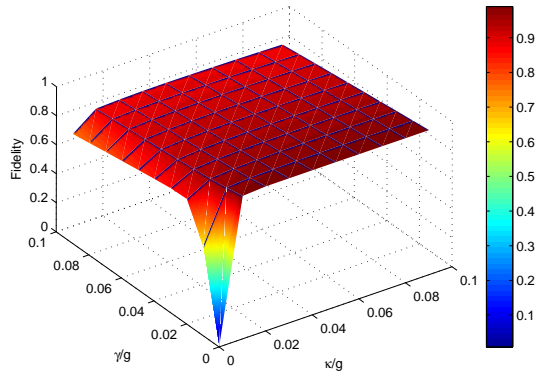}\\
  \caption{Fidelity of the scheme based on spontaneous emission and cavity decay via adding feedback operations $\{U_{c_{L1}} = I$, $U_{c_{R1}} = \exp(i\pi\sigma_{x1}/2)$, $U_{c_{L2}} = I$, $U_{c_{R1}}= I\}$ at the time 20000/\emph{g}.
   Parameters chosen here are the same to Fig.~\ref{f005}.}\label{f006}
\end{figure}

\section{Discussion}\label{s004}

The main method used here is the effective operator method proposed in Refs.~\cite{011,038}. And the main idea of the presented scheme is to leverage dissipative to build effective decay channels and construct the steady state subspace which contains the desired state. Then, effective Hamiltonian is designed to make sure the desired state being its dark state while others evolve out of the subspace. To see clearly the role of each dissipative factors, we first consider the system without cavity decay, and then consider it without spontaneous emission. Numerical simulation shows that both spontaneous emission and cavity decay could be used as resources for high-dimensional entanglement preparation. The only drawback is that cavity-decay-based case is not as ideal as spontaneous-emission-based case. The difference of the effective dissipative channels between these two cases is the chief cause that gives rise to this phenomenon although the steady state subspaces are the same. Therefore, feedback control is added to improve the performance of the cavity-decay-based case. Interestingly, the conditions to obtain the effective dissipative channels in Eq.~(\ref{e009}) and Eq.~(\ref{e013})
can be satisfied at the same time, it is thus possible to treat both spontaneous emission and cavity decay as resources simultaneously.
For coupled cavity
system, coupling strength $g$, cavity decay rate $\kappa$ and the spontaneous emission rate $\gamma$ are stationary, however, we can adjust the parameters $\Omega$, $\omega$, $\Delta$ and $\delta$ to achieve the desired state with higher fidelity. Another important parameter in coupled cavities is the photon-hopping strength \emph{J}, we plot the fidelity of state $|T_{1}\rangle$ versus time under the parameters $(g,\kappa,\gamma)/2\pi\sim$(750, 2.65, 3.5) MHz extracted from an experiment~\cite{049} without feedback control in Fig.~7, from which we can learn that our scheme has great robustness to the variety of $J$. In Fig.~8, without feedback control, we plot the fidelity of the desired state versus time with the parameters the same to Fig.~7 and $J = 6g$. Result shows that fidelity is higher than 97.2\%, which exceeds the values in the unitary-dynamics-based schemes~\cite{035,033,034}.
Moreover, Fig.~7 and Fig.~8 demonstrate that the presented scheme is also feasible without feedback control.
\begin{figure}
  % Requires \usepackage{graphicx}
  \includegraphics[scale = 0.5]{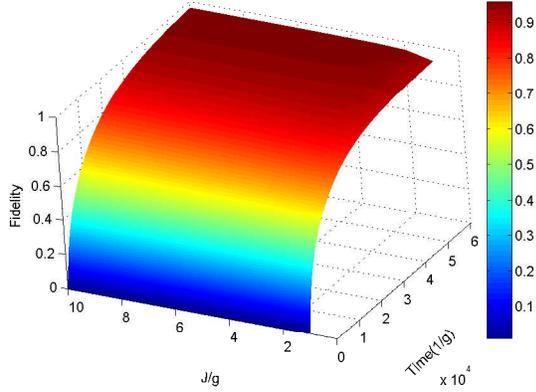}\\
  \caption{Fidelity of $|T_{1}\rangle$ versus $J$ and time without feedback control. The other parameters are chosen as $\Omega = 0.02g$, $\omega = 0.4\Omega$, $\Delta = g$.}\label{f007}
\end{figure}
\begin{figure}
  % Requires \usepackage{graphicx}
  \includegraphics[scale = 0.5]{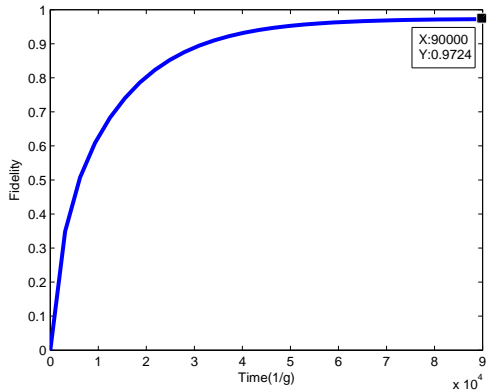}\\
  \caption{Fidelity for generation of three-dimensional entangled state using an experimental cavity parameters without feedback control.}\label{f008}
\end{figure}

\section{Conclusion}\label{s005}
In summary, we have proposed a scheme to prepare a three-dimensional entangled state via using the dissipation. Spontaneous emission and cavity decay are investigated to achieve the desired state, respectively and simultaneously. Moreover, we have investigated the influence of the feedback control for this scheme. Final numerical simulation based on one group of experiment parameters shows that our scheme is feasible under current technology.

\begin{center}{\bf{ACKNOWLEDGMENT}}
\end{center}
This work was supported by the National Natural Science Foundation
of China under Grant Nos. 61068001 and 11264042. X. Q. Shao was also supported by Fundamental Research Funds
for the Central Universities under Grant No. 12SSXM001,
National Natural Science Foundation of China under Grants
No. 11204028, and the Government of China through CSC.

%the Program for
%Chun Miao Excellent Talents of Jilin Provincial Department of
%Education under Grant No. 201316; and the Talent Program of
%Yanbian University of China under Grant No. 950010001.

\end{document}